\documentclass[trackchanges,twocolumn]{aastex701}

\begin{document}
\title{High-order Paschen emission from the quiet-Sun off-limb chromosphere}

\author[0009-0005-8311-1703,sname=Yu,gname=Haocheng]{Haocheng~Yu}
\affiliation{National Astronomical Observatory of Japan, 2-21-1 Osawa, Mitaka, Tokyo 181-8588, Japan}\affiliation{School of Astronomy and Space Science, Nanjing University, 163 Xianlin Road, Nanjing 210023, PR China}
\email{dz21260012@smail.nju.edu.cn}
\author[orcid=0000-0002-5054-8782,sname='Katsukawa']{Yukio~Katsukawa} \affiliation{National Astronomical Observatory of Japan, 2-21-1 Osawa, Mitaka, Tokyo 181-8588, Japan}\affiliation{Department of Astronomy, The University of Tokyo, 7-3-1, Hongo, Bunkyo-ku, Tokyo 113-0033, Japan}\affiliation{Department of Astronomical Science, The Graduate University for Advanced Studies (SOKENDAI), 2-21-1 Osawa, Mitaka, Tokyo 181-8588, Japan}\email{yukio.katsukawa@nao.ac.jp}
\author[]{Mingde~Ding} 
\affiliation{School of Astronomy and Space Science, Nanjing University, 163 Xianlin Road, Nanjing 210023, PR China}
\affiliation{Key Laboratory for Modern Astronomy and Astrophysics, Nanjing 210023, 163 Xianlin Road, PR China}
\email{dmd@nju.edu.cn}

\author[orcid=0000-0002-1043-9944,sname='Matsumoto']{Takuma~Matsumoto} \affiliation{Centre for Integrated Data Science, Institute for Space-Earth Environmental Research, Nagoya University, Furocho, Chikusa-ku, Nagoya, Aichi 464-8601, Japan}\email{takuma.matsumoto@gmail.com}

\author[orcid=0000-0002-3418-8449,sname='Solanki']{Sami~K.~Solanki} \affiliation{Max-Planck-Institut für Sonnensystemforschung, Justus-von-Liebig-Weg 3, 37077 Göttingen, Germany}\email{solanki@mps.mpg.de}

\author[orcid=0000-0002-2055-441X,sname='Blanco~Rodríguez']{Julian~Blanco~Rodríguez} \affiliation{Universitat de Valencia Catedrático José Beltrán 2, E-46980 Paterna-Valencia, Spain}\affiliation{Spanish Space Solar Physics Consortium}\email{julian.blanco@uv.es}

\author[orcid=0000-0001-8829-1938,sname='Orozco~Suárez']{David~Orozco~Suárez} \affiliation{Instituto de Astrofísica de Andalucía, CSIC, Glorieta de la Astronomía s/n, 18008 Granada, Spain}\affiliation{Spanish Space Solar Physics Consortium}\email{orozco@iaa.es}

\author[orcid=0000-0001-5616-2808,sname='Kubo']{Masahito~Kubo} \affiliation{National Astronomical Observatory of Japan, 2-21-1 Osawa, Mitaka, Tokyo 181-8588, Japan}\affiliation{Department of Astronomical Science, The Graduate University for Advanced Studies (SOKENDAI), 2-21-1 Osawa, Mitaka, Tokyo 181-8588, Japan}\email{masahito.kubo@nao.ac.jp}


\author[orcid=0000-0003-1459-7074,sname='Lagg']{Andreas~Lagg} \affiliation{Max-Planck-Institut für Sonnensystemforschung, Justus-von-Liebig-Weg 3, 37077 Göttingen, Germany}\email{lagg@mps.mpg.de}

\author[orcid=0000-0002-9972-9840,sname='Gandorfer']{Achim~Gandorfer} \affiliation{Max-Planck-Institut für Sonnensystemforschung, Justus-von-Liebig-Weg 3, 37077 Göttingen, Germany}\email{gandorfer@mps.mpg.de}

\author[orcid=0000-0002-3387-026X,sname='del~Toro~Iniesta']{Jose~Carlos~del~Toro~Iniesta} \affiliation{Instituto de Astrofísica de Andalucía, CSIC, Glorieta de la Astronomía s/n, 18008 Granada, Spain}\affiliation{Spanish Space Solar Physics Consortium}\email{jti@iaa.es}

\author[orcid=0000-0002-0787-8954,sname='Bernasconi']{Pietro~Bernasconi} \affiliation{Johns Hopkins University Applied Physics Laboratory, 11100 Johns Hopkins Road, Laurel, Maryland, USA}\email{pietro.bernasconi@jhuapl.edu}
\author[sname='Berkefeld']{Thomas~Berkefeld} \affiliation{Institut für Sonnenphysik (KIS), Georges-Köhler-Allee 401a, 79110 Freiburg, Germany}\email{thomas.berkefeld@leibniz-kis.de}
\author[orcid=0009-0009-4425-599X,sname='Feller']{Alex~Feller} \affiliation{Max-Planck-Institut für Sonnensystemforschung, Justus-von-Liebig-Weg 3, 37077 Göttingen, Germany}\email{feller@mps.mpg.de}
\author[orcid=0000-0001-6317-4380,sname='Riethmüller']{Tino~L.~Riethmüller} \affiliation{Max-Planck-Institut für Sonnensystemforschung, Justus-von-Liebig-Weg 3, 37077 Göttingen, Germany}\email{riethmueller@mps.mpg.de}

\author[orcid=0000-0001-9228-3412,sname='Álvarez-Herrero']{Alberto~Álvarez-Herrero} \affiliation{Instituto Nacional de T\'ecnica Aeroespacial (INTA), Ctra. de Ajalvir, km. 4, E-28850 Torrejón de Ardoz, Spain}\affiliation{Spanish Space Solar Physics Consortium}\email{alvareza@inta.es}
\author[orcid=0000-0003-3490-6532,sname='Smitha']{H.~N.~Smitha} \affiliation{Max-Planck-Institut für Sonnensystemforschung, Justus-von-Liebig-Weg 3, 37077 Göttingen, Germany}\email{narayanamurthy@mps.mpg.de}
\author[sname='Grauf']{Bianca~Grauf} \affiliation{Max-Planck-Institut für Sonnensystemforschung, Justus-von-Liebig-Weg 3, 37077 Göttingen, Germany}\email{grauf@mps.mpg.de}
\author[sname='Carpenter']{Michael~Carpenter} \affiliation{Johns Hopkins University Applied Physics Laboratory, 11100 Johns Hopkins Road, Laurel, Maryland, USA}\email{michael.carpenter@jhuapl.edu}
\author[sname='Bell']{Alexander~Bell} \affiliation{Institut für Sonnenphysik (KIS), Georges-Köhler-Allee 401a, 79110 Freiburg, Germany}\email{albe@leibniz-kis.de}
\author[orcid=0000-0001-7764-6895,sname='Martínez~Pillet']{Valentín~Martínez~Pillet} \affiliation{Instituto de Astrofísica de Canarias, Vía Láctea, s/n, E-38205 La Laguna, Spain}\affiliation{Spanish Space Solar Physics Consortium}\email{vmpillet@iac.es}

\author[orcid=0000-0002-7318-3536,sname='Bailén']{Francisco~Javier~Bailén} \affiliation{Instituto de Astrofísica de Andalucía, CSIC, Glorieta de la Astronomía s/n, 18008 Granada, Spain}\affiliation{Spanish Space Solar Physics Consortium}\email{fbailen@iaa.es}
\author[orcid=0000-0003-4319-2009,sname='Castellanos~Durán']{Juan~Sebastián~Castellanos~Durán} \affiliation{Max-Planck-Institut für Sonnensystemforschung, Justus-von-Liebig-Weg 3, 37077 Göttingen, Germany}\email{castellanos@mps.mpg.de}
\author[orcid=0009-0002-6808-5154,sname='Harnes']{Edvarda~Harnes} \affiliation{Max-Planck-Institut für Sonnensystemforschung, Justus-von-Liebig-Weg 3, 37077 Göttingen, Germany}\email{harnes@mps.mpg.de}
\author[orcid=0000-0001-6029-7529,sname='Hölken']{Johannes~Hölken} \affiliation{Max-Planck-Institut für Sonnensystemforschung, Justus-von-Liebig-Weg 3, 37077 Göttingen, Germany}\email{hoelken@mps.mpg.de}
\author[orcid=0000-0003-1409-1145,sname='Iglesias']{Francisco~A.~Iglesias} \affiliation{Max-Planck-Institut für Sonnensystemforschung, Justus-von-Liebig-Weg 3, 37077 Göttingen, Germany}\affiliation{Grupo de Estudios en Heliofísica de Mendoza, CONICET, Universidad de Mendoza, Boulogne sur Mer 683, 5500 Mendoza, Argentina}\email{iglesias@mps.mpg.de}
\author[orcid=0000-0002-4669-5376,sname='Ishikawa']{Ryohtaroh~T.~Ishikawa} \affiliation{National Institute for Fusion Science, 322-6 Oroshi-cho, Toki City 509-5292, Japan}\email{ishikawa.ryohtaro@nifs.ac.jp}
\author[orcid=0000-0001-7452-0656,sname='Kawabata']{Yusuke~Kawabata} \affiliation{National Astronomical Observatory of Japan, 2-21-1 Osawa, Mitaka, Tokyo 181-8588, Japan}\email{kawabata.yusuke@nao.ac.jp}
\author[orcid=0000-0002-7044-6281,sname='Oba']{Takayoshi~Oba} \affiliation{Advanced Research Center for Space Science and Technology, Institute of Science and Engineering, Kanazawa University, Kakuma-machi, Kanazawa, Ishikawa 920-1192, Japan}\affiliation{Max-Planck-Institut für Sonnensystemforschung, Justus-von-Liebig-Weg 3, 37077 Göttingen, Germany}\email{oba@mps.mpg.de}
\author[orcid=0000-0003-0175-6232,sname='Siu-Tapia']{Azaymi~L.~Siu-Tapia} \affiliation{Instituto de Astrofísica de Andalucía, CSIC, Glorieta de la Astronomía s/n, 18008 Granada, Spain}\affiliation{Spanish Space Solar Physics Consortium}\email{siu@iaa.es}
\author[orcid=0000-0003-1483-4535,sname='Strecker']{Hanna~Strecker} \affiliation{Instituto de Astrofísica de Andalucía, CSIC, Glorieta de la Astronomía s/n, 18008 Granada, Spain}\affiliation{Spanish Space Solar Physics Consortium}\email{streckerh@iaa.es}
\author[orcid=0000-0003-1971-5551,sname='Vukadinović']{Dušan~Vukadinović} \affiliation{Institut für Physik, Universität Graz, Universitätsplatz 5, 8010 Graz, Austria}\affiliation{Max-Planck-Institut für Sonnensystemforschung, Justus-von-Liebig-Weg 3, 37077 Göttingen, Germany}\email{dusan.vukadinovic@uni-graz.at}

\author[orcid=0000-0001-5686-3081,sname='Hara']{Hirohisa~Hara} \affiliation{National Astronomical Observatory of Japan, 2-21-1 Osawa, Mitaka, Tokyo 181-8588, Japan}\email{hirohisa.hara@nao.ac.jp}
\author[orcid=0000-0003-4764-6856,sname='Shimizu']{Toshifumi~Shimizu} \affiliation{Department of Earth and Planetary Science, The University of Tokyo, 7-3-1, Hongo, Bunkyo-ku, Tokyo 113-0033, Japan}\affiliation{Institute of Space and Astronautical Science, Japan Aerospace Exploration Agency, 3-1-1, Yoshinodai, Chuo-ku, Sagamihara, Kanagawa 252-5210, Japan}\email{shimizu.toshifumi@isas.jaxa.jp}

\begin{abstract}

We report the detection of high-order hydrogen Paschen emission lines (Pa~15, Pa~16, and Pa~17) in the quiet-Sun chromosphere off the solar limb using the Chromospheric Infrared SpectroPolarimeter (SCIP) on board the {\sc Sunrise~iii} balloon telescope. These lines reveal thread-like structures resembling spicules and exhibit systematically smaller Doppler velocities than Ca~II~854.2~nm, suggesting that they are optically thinner and more affected by line-of-sight averaging, especially near the limb. Non-LTE radiative transfer synthesis using the spherically symmetric one-dimensional code \texttt{rhsphere} reproduces the overall spectral properties. The observed ratios among three Paschen lines show systematic deviations from synthetic and theoretical results, suggesting that additional physical effects may influence the formation of high-order Paschen lines. The study demonstrates the potential of high-order Paschen lines as a new diagnostic of optically thin plasma in the off-limb chromosphere.

\end{abstract}

\section{Introduction}

High-order hydrogen Paschen lines are intrinsically weak in the solar atmosphere due to the low population of high principal quantum levels, resulting in low opacity and weak line-to-continuum contrast relative to lower-order Balmer and Paschen lines \citep{2003Rutten}. Consequently, these lines are difficult to detect on the solar disk, where the strong photospheric background dominates. Previous observations have therefore mainly focused on active structures such as prominences and jets, where enhanced emission conditions prevail \citep{1986Engvold,1987Foukal,2014Anan}.

Off-limb observations provide a more favorable geometry, as the line of sight is largely free from photospheric continuum contamination. Under such conditions, hydrogen level populations may approach the Case B recombination regime \citep{1938Baker}, in which Lyman transitions are optically thick while higher-series lines remain optically thin and are efficiently produced through recombination cascades. Calculations by \citet{1987Hummer} predict that high-order Paschen lines can reach appreciable emissivity under coronal or prominence-like conditions.

In this Letter, we present observations of high-order Paschen emission (Pa~15, Pa~16, and Pa~17) in a quiet-Sun off-limb region obtained with {\sc Sunrise~iii}/SCIP. Unlike previous studies, the observations target a non-active region.

\section{SCIP observations}
\label{SCIP observation}

In July 2024, the third science flight of the {\sc Sunrise} telescope was successfully carried out \citep{2025Korpi-Lagg, 2026Solanki}. Sunrise III builds on the earlier version of Sunrise \citep{2011Barthol} that successfully flew in 2010 \citep{2010Solanki} and 2013 \citep{2017Solanki}. SCIP is one of the two slit-based spectropolarimeters on board, operating in the near-infrared passband \citep{2026Katsukawa}, the other being SUSI \citep{2025Feller} that operates in the near UV. The first spectral channel of SCIP includes two lines of the Ca\, II infrared triplet at 849.8 nm and 854.2 nm, together with several hydrogen Paschen lines, namely Pa\,15 (854.5 nm), Pa\,16 (850.2 nm), and Pa\,17 (846.7 nm).

From 22:13 UT to 23:09 UT on 13 July 2024, SCIP scanned the solar limb at approximately $(415'', -830'')$ in helioprojective coordinates. Each slit image provides spatial information along the slit and spectral information along the wavelength axis. The calibrated spectra consist of 2150 wavelength pixels with a sampling of 0.0395 \AA. The spatial sampling along the slit is $0\farcs094$ per pixel, with 640 pixels ($60''$) covering the slit length. The effective integration time at each slit position was 10.24 s. The data were processed using dark-current subtraction, skew and wavelength calibration, flat-field correction, and polarimetric calibration (see \citealt{2026Solanki} for more details). By stacking multiple slit positions, we constructed a three-dimensional cube containing two spatial dimensions and one spectral dimension. To suppress oscillatory patterns and stripe-like artifacts in the spectra, the data were smoothed along the spectral direction using a Savitzky–Golay filter \citep{1964Savitzky}.
    
\subsection{Treatment of the off-limb background}

The stray light is non-negligible off the solar limb. The observed spectra contain a spectral background that includes instrumental stray light and genuine solar background radiation. The instrumental stray light is expected to be dominated by leakage from the bright solar disk. Since Paschen lines are hardly visible in the disk spectrum, this stray-light component should appear as a spectrally smooth continuum background in the vicinity of the Paschen lines. We determine this background locally at each spatial position by fitting each Paschen-line profile with a Gaussian component plus a linear background (see Sect.~\ref{Doppler velocity}). Over the narrow fitting interval, the derived Gaussian intensity represents the line emission above this local background, and the linear term represents the slowly varying local background beneath the Paschen line, including the contribution of stray light and the solar continuum radiation.

This procedure is a local spectral-background subtraction rather than a spatial deconvolution. The latter would require a well-characterized effective point-spread function, including the slit response and extended scattering wings, and would therefore need careful instrument calibration. The absence of Paschen lines in the intense disk radiation justifies local spectral-background subtraction in the off-limb spectral data. To assess the influence of our treatment, we repeated the analysis using an alternative height-dependent stray-light correction based on \citet{2011Beck}. The mean relative differences in the integrated line intensities between the two treatments are less than 2\% for all three Paschen lines over the pixels with reliable Gaussian fits.

\section{Results}
\label{Result}

\begin{figure*}[htbp]
    \centering
    \includegraphics[width=\linewidth]{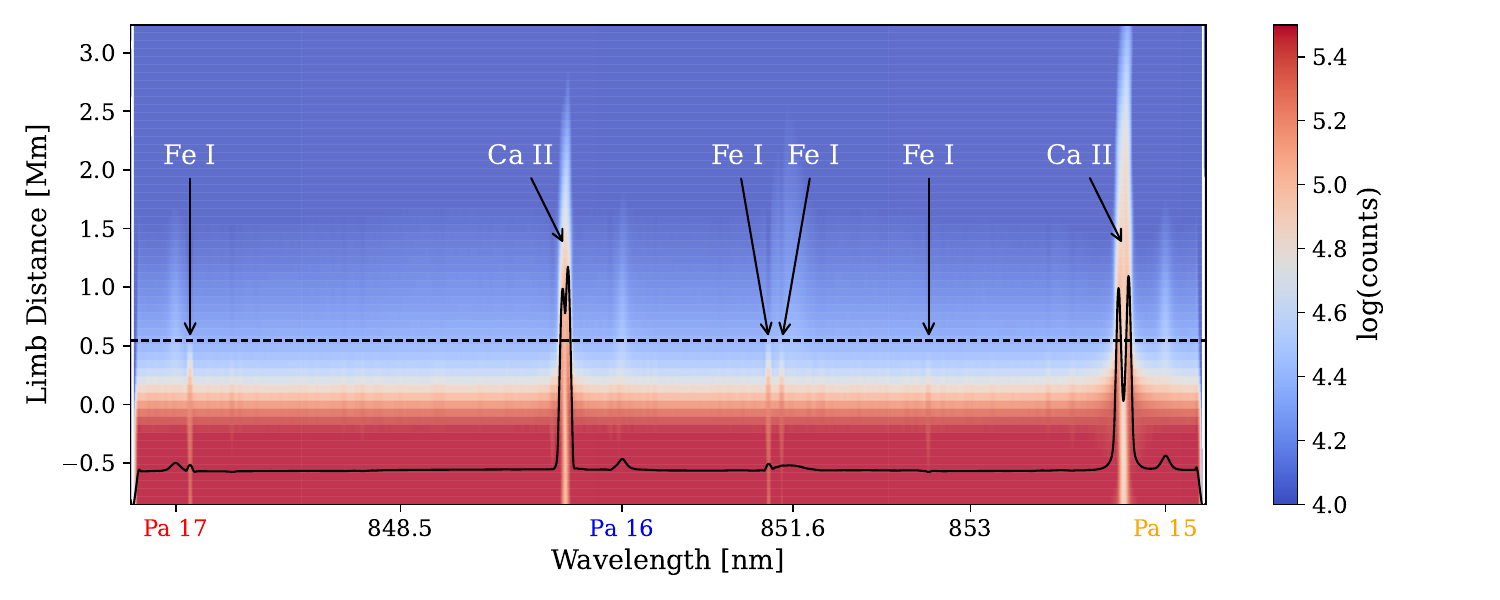}
    \caption{Wavelength–spatial map of the slit at 22:18 UT. The positions of Pa 15, Pa 16, and Pa 17 are indicated by orange, blue, and red marks along the abscissa, respectively. Four neutral iron lines (Fe~I) and two calcium lines (Ca~II) are indicated by arrows. The representative spectrum shown in the figure (solid black line) is extracted along the black dashed line.}
    \label{wave_space}
\end{figure*}

Fig.~\ref{wave_space} displays the wavelength–spatial map of a representative slit, together with a representative line profile obtained off the solar limb. The Ca II lines at 849.8 nm and 854.2 nm are the strongest lines in this spectral range, which change from absorption on the solar disk to emission off the limb. Three Paschen lines are labeled on the abscissa. Within the limb in Fig.~\ref{wave_space}, these three high-order Paschen lines can hardly be seen, because they are blended with weak photospheric lines. However, outside the solar limb, these Paschen lines show apparent single peaks, while the blending photospheric lines disappear. We note that there is another strong and broad emission signal at about 851.6 nm off the limb, which is an instrumental artifact due to internal stray light.

\subsection{Off-limb Threads}

\begin{figure*}
    \centering
    \includegraphics[width=\linewidth]{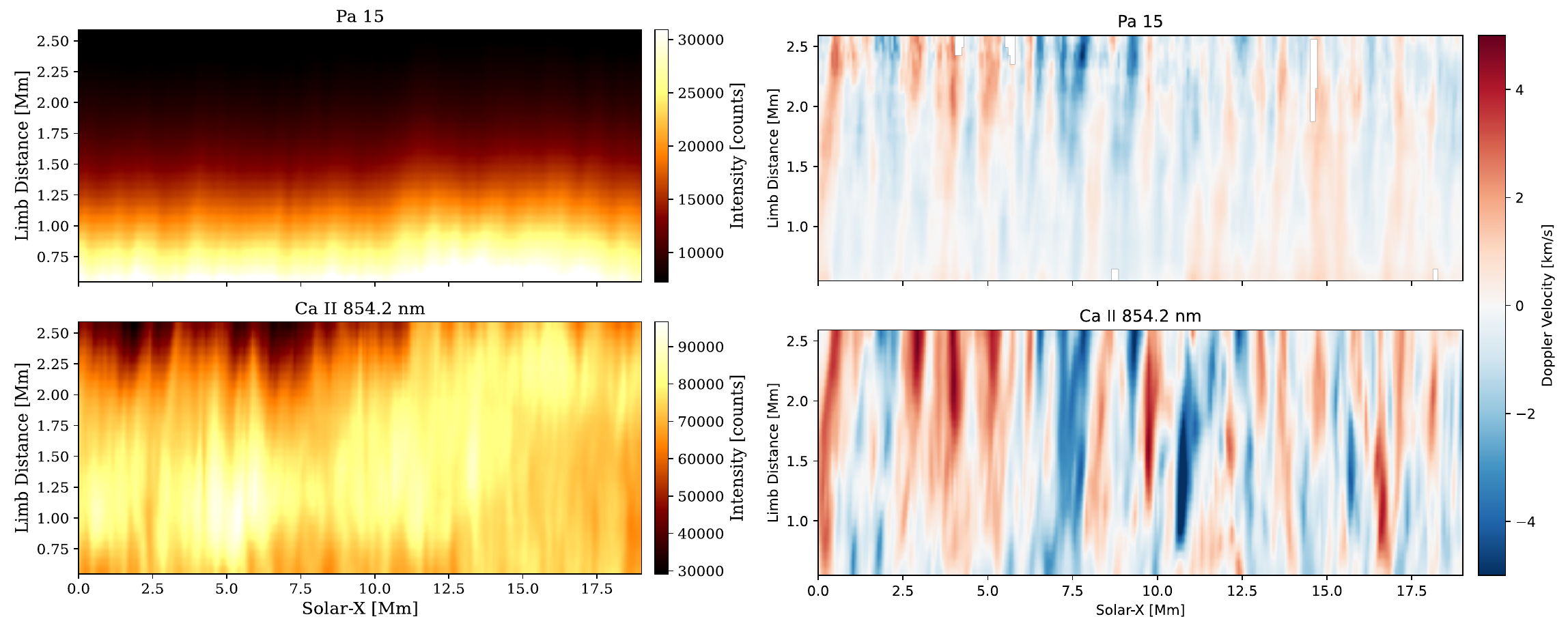}
    \caption{Left: Intensity maps of Pa~15 line together with the Ca~II~854.2~nm line. All maps are constructed from intensities measured at the corresponding line center. Right: Dopplergrams of the two lines. White patches in Pa~15 are pixels that failed Gaussian fitting.}
    \label{combine}
\end{figure*}

The observation of the chromospheric lines is dominated by omnipresent jet-like structures, which are commonly referred to as spicules when observed off the solar limb \citep{2012Tsiropoula}. Since the slit direction was approximately perpendicular to the solar limb, we constructed the map by rotating each slit by only about $3^\circ$. The left part of Fig~\ref{combine} displays a two-dimensional image of the solar limb observation of the Pa~15 line, together with that of the Ca II 854.2 nm line. Although we do not show Pa~16 and Pa~17, the images among the three Paschen lines are nearly the same. All the Paschen lines show thread-like structures, resembling spicules observed in the chromospheric lines \citep{2007Pontieu,2010Judge}. The fine structures in the Paschen lines are detected only between 0.75 Mm and 1.75 Mm, while below 0.75 Mm they are submerged in the continuum background and cannot be reliably identified. The image of the Ca II lines shows fine threads extending above 2.5 Mm, with a larger height scale and clearer structures. Although the threads in Paschen lines are much weaker, the structures revealed by these lines are quite distinct from the continuum image at 850.37 nm, which shows no thread-like structures.

\subsection{Doppler velocity}
\label{Doppler velocity}
Off the solar limb, the Paschen lines perform as a single peak emission; thus, we fit each Paschen line with a Gaussian function.
\begin{equation}
    I(\lambda) = A\mathrm{exp}(-\frac{(\lambda-\lambda_0)^2}{2\sigma^2}) + B(\lambda-\lambda_0)+C.
\label{gaussian fitting}
\end{equation}

Here, a linear background is used to fit the Ca II wings, especially for Pa 15. The Doppler velocity of the spectral line is determined as $(\lambda_0-\lambda_c)\cdot c/\lambda_c$, where $\lambda_c$ is the static wavelength and $c$ is the speed of light. For the Ca II 849.8 nm and 854.2 nm lines, we determine the Doppler velocity using the center-of-gravity method, which accounts for emission profiles with core absorption as well as those with a single emission peak.

\begin{figure}
    \centering
    \includegraphics[width=\linewidth]{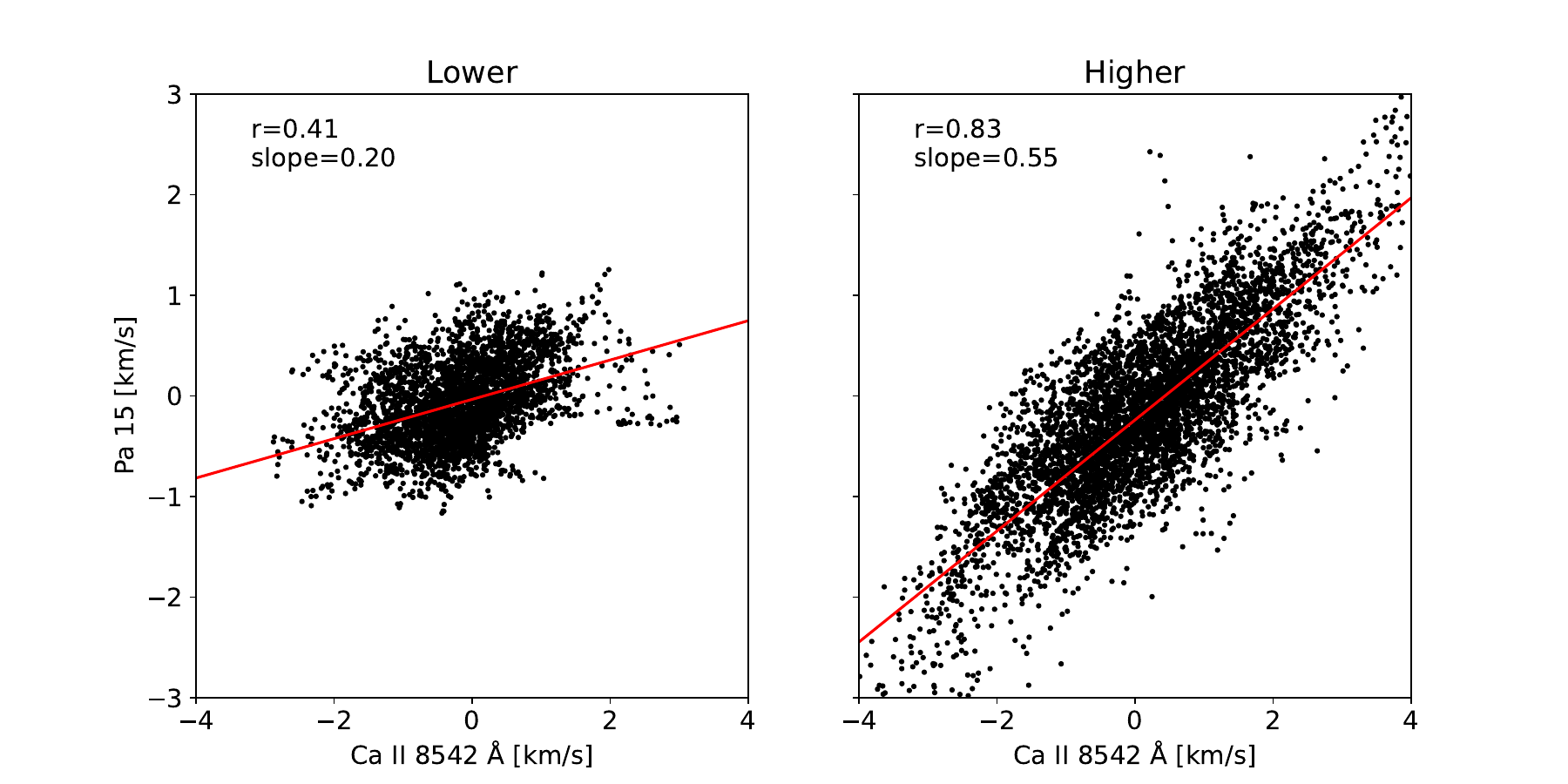}
    \caption{Scatter plots of the Doppler velocity between Pa 15 and Ca II 854.2 nm at smaller (0.5 - 1.4 Mm) and greater heights above the limb (1.4 - 2.6 Mm).}
    \label{Correlation}
\end{figure}

The right part of Fig.~\ref{combine} shows the Doppler velocity of both Ca II 854.2 nm and Pa 15 lines. There are other emission signals located in the blue and red wings of Pa 17 and Pa 16, which significantly influence the accuracy of Gaussian fitting; thus, we only show Pa 15. Similar to the intensity maps, the Dopplergrams of the Ca and Paschen lines have thread-like structures, that are roughly perpendicular to the solar limb. The Dopplergram of Pa 15 shows clear differences from that of Ca II, especially near the limb. It appears as if the velocities from Ca II are larger than from Pa 15. Therefore, we plot the velocity scatters between Ca II 854.2 nm and Pa 15 at different heights in Fig.~\ref{Correlation}. In the lower region (0.5 - 1.4 Mm), the velocity of Pa 15 spreads between $\pm1$ km/s, while the velocity of Ca II ranges from -3 to 3 km/s. The slope between the two lines is only 0.2, indicating that the velocities of Ca II are five times larger than those of Pa 15. Also, the correlation between the two lines is small. In the higher region (1.4 - 2.6 Mm), both Pa 15 and Ca II exhibit a broader velocity distribution, the slope increases to 0.55 and the correlation coefficient to 0.83.

This increase can be understood in terms of optical depth effects. Pa 15 is optically thinner than Ca II. Near the limb, there are more emitting structures along the line-of-sight (LOS), so multiple velocity components contribute to the Pa 15 emission, leading to significant averaging and velocities closer to zero. In contrast, the optically thicker Ca II line primarily samples the foreground layers of the emitting structures. At larger heights, the emitting structures become sparser, reducing the LOS averaging effect and increasing the slope. Meanwhile, the reduced averaging may also lead to a broader velocity range in the higher region. These results support the interpretation that the Paschen lines are optically thinner than Ca II and are therefore more affected by LOS averaging.

\section{Off-limb synthesis}
\label{Discussion}

\begin{figure}
    \centering
    \includegraphics[width=\linewidth]{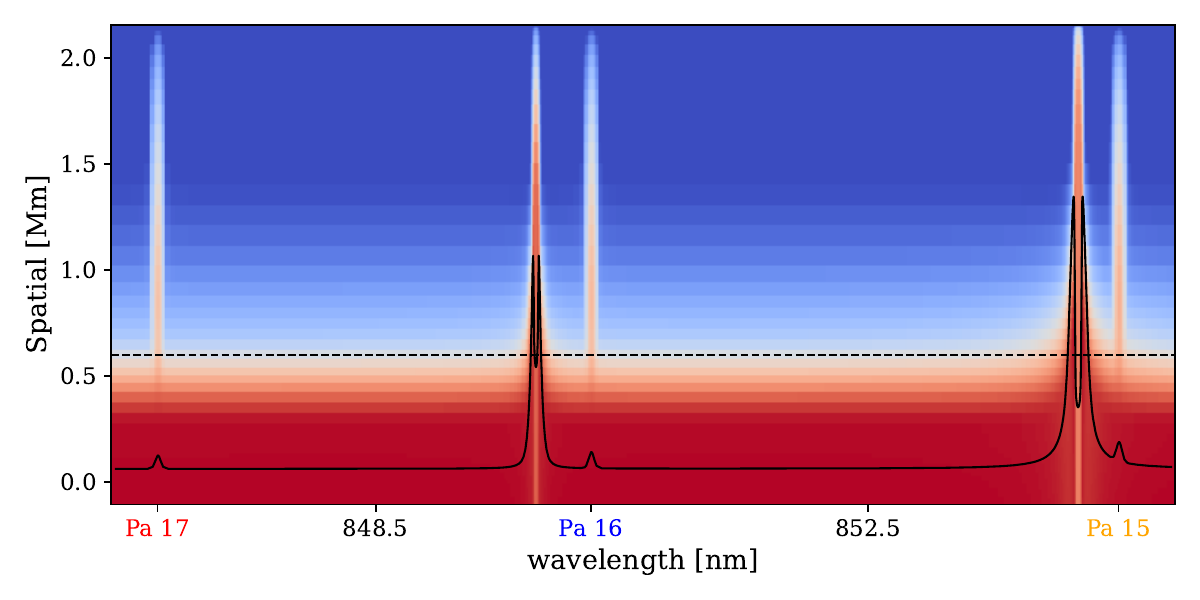}
    \caption{The same as Fig.\ref{wave_space} but synthesized in a spherically symmetric atmosphere.}
    \label{simulation}
\end{figure}

To compare with the observation, we used a code for radiative transfer in spherical geometry, \verb"rhsphere" \citep{2001Uitenbroek}, to synthesize the off-limb spectrum. We use a Hydrogen model with 19 bound levels in synthesis. Both Hydrogen and Calcium lines are considered in non-LTE and partial redistribution (PRD). The atmospheric model employed is FALC \citep{1993Fontenla}. The emergent spectra are computed at very small heliocentric angles, corresponding to a set of concentric rays with increasing radii, which can be converted into off-limb distances. The resulting wavelength–spatial map is shown in Fig.~\ref{simulation}. Away from the solar limb, the Paschen lines appear as single-peaked emission profiles, similar to the observed features (see Fig.~\ref{wave_space}). We note that \verb"rhsphere" is a 1D spherically symmetric model and cannot reproduce the observed 2D structures in Fig~\ref{combine}.

\begin{figure}[htbp]
    \centering
    \includegraphics[width=\linewidth]{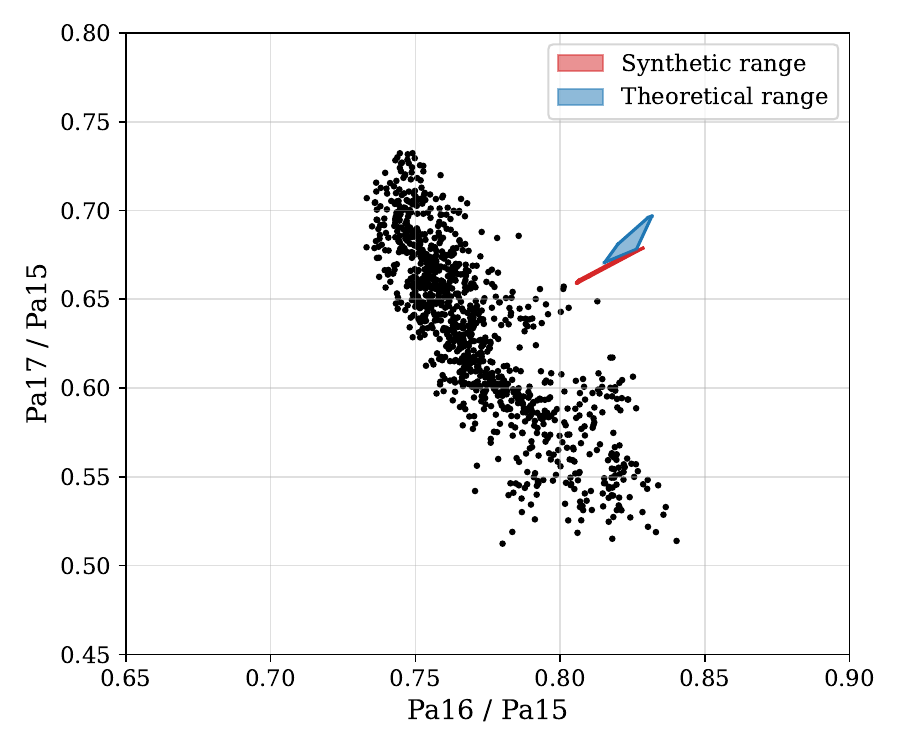}
    \caption{Scatters between intensity ratios of Pa~16/Pa~15 and Pa~17/Pa~15. Results from spherical synthesis and theoretical calculations (Case B) are plotted as red and blue shaded regions, respectively.}
    \label{ratio}
\end{figure}

Fig~\ref{ratio} shows a scatter plot of the intensity ratio Pa 16/Pa 15 and Pa 17/Pa 15 of the observation, in which the line intensity is calculated by $A\sigma\sqrt{2\pi}$ in Equation~\ref{gaussian fitting}. The intensity ratios among Pa 15, 16, and 17 show a spread, with the value Pa~16/Pa~15 ranging from 0.73 to 0.83, and Pa~17/Pa~15 ranging from 0.53 to 0.74. In comparison, the synthetic ratios are weakly dependent on height and cluster around 1:0.81:0.67 (Pa~15:Pa~16:Pa~17). We also plot the theoretical range given by \cite{1987Hummer}, covering temperatures from $5000$ to $30000\ \mathrm{K}$ and electron density from $10^{9}$ to $10^{10}\ \mathrm{cm^{-3}}$, close to the synthetic results. Under the Case B assumption, variations in temperature and density produce a positive correlation between Pa~16/Pa~15 and Pa~17/Pa~15. In contrast, the observed ratios exhibit an apparent anti-correlation, indicating that the formation of these lines cannot be fully explained by simple Case B recombination calculations. The \verb"rhsphere" synthesis roughly reproduces the observed ratio, but noticeable discrepancies remain. These differences suggest that additional physical effects, not included in the simplified one-dimensional spherical model (such as multidimensional plasma structure and inhomogeneity), may influence the formation of the off-limb Paschen emission.

\section{Conclusion}
\label{Conclusion}

In this work, we report the clear detection of high-order Paschen lines (Pa~15, Pa~16, and Pa~17) in the quiet-Sun chromosphere off the solar limb using {\sc Sunrise~iii}/SCIP observations. Maps made with these lines exhibit spicule-like structures and systematically smaller Doppler velocities than Ca~II~854.2~nm. These properties are likely because they are optically thinner and hence more affected by LOS averaging.

Radiative transfer synthesis reproduces the overall spectral characteristics. However, the observed line ratios show significant dispersion and deviate from one-dimensional model predictions and Case B theoretical calculations, suggesting that multi-dimensional and inhomogeneous plasma conditions may contribute to the line formation.

These results demonstrate that high-order Paschen lines serve as a new probe of optically thin plasma in the off-limb chromosphere. In addition to Paschen lines, future spectroscopic observations of other lines near and beyond the solar limb (e.g., \citealt{CastellanosDuran2026ApJ}) may provide complementary diagnostics of fine structures and help to further constrain the physical conditions of the off-limb solar atmosphere.

\begin{acknowledgments}

We thank C. Quintero Noda and Han Uitenbroek for their kind guidance on using the \verb"rhsphere" code. 

Sunrise III is supported by funding from the Max-Planck-Förderstiftung (Max Planck Foundation), NASA under Grants \#80NSSC18K0934 and \#80NSSC24M0024 ("Heliophysics Low Cost Access to Space" program), and the ISAS/JAXA Small Mission-of-Opportunity program and JSPS KAKENHI Grant Numbers JP18H05234 and JP23K25916. This research has received financial support from the European Union’s Horizon 2020 research and innovation program under grant agreement No. 824135 (SOLARNET). It has also been funded by the Deutsches Zentrum für Luft- und Raumfahrt e.V. (DLR, grant no. 50 OO 1608). The Spanish contributions have been funded by the Spanish MCIN/AEI/10.13039/501100011033 under projects RTI2018-096886-B-C5, PID2021-125325OB-C5, and PID2024-156066OB-C5, and from "Center of Excellence Severo Ochoa" awards to IAA-CSIC (SEV-2017-0709, CEX2021-001131-S), all co-funded by "ERDF A way of making Europe".

\end{acknowledgments}

\bibliography{sample701}{}
\bibliographystyle{aasjournalv7}

\end{document}